\documentclass{aa501}
\usepackage{graphicx}

\newcommand{\seco}{^{\prime\prime}}
\renewcommand{\min}{^{\prime}}

\newcommand{\AAS}[2]{{\em A\&AS,\/} {\bf #1}, #2}
\newcommand{\AaA}[2]{{\em A\&A,\/} {\bf #1}, #2}
\newcommand{\AJ}[2]{{\em AJ,\/} {\bf #1}, #2}
\newcommand{\PASP}[2]{{{\em PASP,\/} {\bf #1}, #2}}
\newcommand{\SPIE}[2]{{{\em SPIE,\/} {\bf #1}, #2}}
\def \hang {\hangindent\parindent}
\newcommand{\myitem}{\smallskip\par\noindent\hang}

\begin{document}

\title{NICS: The TNG Near Infrared Camera Spectrometer
\thanks{Based on observations taken at TNG, La Palma.}
}

\author{ 
C.~Baffa\inst{1}\and G.~Comoretto\inst{1}\and S.~Gennari\inst{1}\and F.~Lisi\inst{1}\and 
E.~Oliva\inst{1,4}\and V.~Biliotti\inst{1}\and
A.~Checcucci\inst{1}\and  
V.~Gavrioussev\inst{2}\and 
E.~Giani\inst{1}\and 
F.~Ghinassi\inst{4}\and 
L.K.~Hunt\inst{2}\and
R.~Maiolino\inst{1}\and F.~Mannucci\inst{2}\and G.~Marcucci\inst{3}\and
M.~Sozzi\inst{2}\and 
P.~Stefanini\inst{1}\and
L.Testi\inst{1}
} 
\offprints{C. Baffa}

\institute{
Osservatorio Astrofisico di Arcetri, Largo E. Fermi 5, 
Firenze, Italy
\and
Centro per l'Astronomia Infrarossa e lo Studio del Mezzo
Interstellare--CNR, Largo E. Fermi 5, Firenze, Italy
\and
Dipartimento di Astronomia e Scienze dello Spazio, Universit\`a
di Firenze, Largo E. Fermi 5, Firenze, Italy
\and
Centro Galileo Galilei, Santa Cruz De La Palma, TF, La Palma, Spain.
}

\date{Received..........; Accepted..........}

\abstract{
NICS (acronym for Near Infrared Camera Spectrometer) is the near-infrared cooled
camera--spectrometer that has been developed by the Arcetri Infrared Group
at the Arcetri Astrophysical Observatory, in collaboration with the
CAISMI-CNR for the TNG (the Italian National Telescope Galileo at La Palma,
Canary Islands, Spain).
As NICS is in its scientific commissioning phase, we report
its observing capabilities in the near--infrared bands at the
TNG, along with the measured performance and the limiting
magnitudes. We also describe some technical details
of the project, such as cryogenics, mechanics, and the system which
executes data acquisition and control, along with the related software.
\bigskip
}
\maketitle

\section{Introduction}

The 3.5m Italian National Telescope Galileo (TNG) (Barbieri 1995),
under operation on La Palma (Canary Islands), included a general
purpose near--infrared camera/spectrometer in its Instrument
Plan for first--light operations (Fusi Pecci et al. 1992, Fusi Pecci et
al. 1994). The telescope itself is not optimized for the thermal infrared
bands, so NICS was designed to operate at near--infrared wavelengths,
from 0.95 $\mu$m up to 2.50~$\mu$m, avoiding the spectral range where
the ambient blackbody radiation could degrade the signal--to--noise ratio
of the observations.  Moreover, this spectral range is conveniently
covered by HgCdTe (or MCT for Mercury--Cadmium--Telluride) large format
focal--plane array detectors currently available, which, at present,
offer the best combination of quality and low read-noise. NICS was
the only infrared instrument for the first light of the TNG; as a
consequence, we decided to incorporate a sufficient degree of operation
flexibility by adopting a collimator/camera optical scheme along with a
good pupil image where a number of analyzers (filters, grisms, polarizers)
can be easily accommodated. This configuration allows a large
number of photometric and spectroscopic observing modes, and the observer
can switch rapidly between different modes in remote operation, for
instance, adapting the observations to the seeing conditions of the
night.

\section{Observing modes}
The instrument is provided with the following imaging and spectroscopic
observing modes:

\begin{description} 

\item{--}  
wide-field imaging with a plate scale of $0.25 \seco$/pixel and a total
field, as projected on the sky, of more than $4\min \times 4\min$; 
wide-- and narrow--band filters are available for photometry and
in--line imaging; 

\item{--}  
small-field imaging with a plate scale of $0.13 \seco$/pixel ($\sim
2\min\ \times\ 2\min$ field of view), for better sampling under excellent
seeing conditions;

\item{--}  
medium- to low-dispersion long--slit ($4 \min$ slit) grism spectroscopy with a
resolving power between 300 and 1300; 

\item{--} 
very low-dispersion long--slit ($4 \min$ slit) spectroscopy with a
resolving power $\approx 50$, by means of an Amici prism;

\item{--} 
imaging polarimetry for both wide-- and small--field imaging mode.
Polarimetry imaging is performed on only 1/4 of the field of view, but
simultaneously on four directions of polarization angle (0, 45, 90, and
135 degrees), with a clear gain of relative sensitivity.

\item{--} 
spectro--polarimetry with a reduced (25\%) slit length, but with four
directions of polarization angle (0, 45, 90, and 135 degrees) measured
simultaneously.

\end{description}

Three of these modes, low-dispersion long--slit spectroscopy and
simultaneous angle polarimetry/spectro--polarimetry are unique to
NICS, and provide for the first time the possibility to perform
reliably these measurements in the infrared.

\begin{figure}
\centerline{\includegraphics[width=7.3cm]{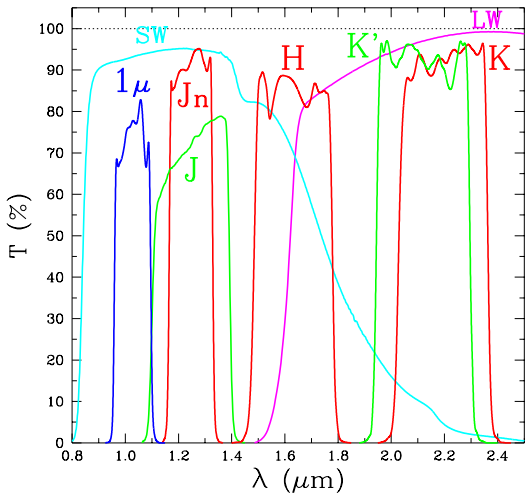}}
\caption{ 
Spectral response and efficiency of NICS wide--band filters.
}
\end{figure}

The simultaneous mode polarimetry is obtained by means of wedged double
Wollaston prisms and a special field stop (Oliva 1997). High
spatial resolution imaging at the diffraction limit of the telescope is
possible by means of an external adaptive optics module (Ragazzoni 1996),
using the re-imaging optics of the adaptive module, which has an f/33
output beam. Two reduced plate scales (about $0.08 \seco$ and $0.04 \seco$
per pixel, respectively) are available in connection with the wide
field and the small-field optics of NICS.

Figure 1 shows the spectral response and the efficiency of the wide band
filters which are presently available; besides the standard filters for J,
H, and K bands, NICS offers the 1$\mu$m filter centered at 1.030 $\mu$m in
correspondence with a fair atmospheric window, the J$_n$ filter as defined by
the Gemini project, the K' filter which cuts 
the K band portion where the thermal emission dominates the
background flux. There is also a band pass filter (labeled SW) for general
purpose observations and pointing and a high--pass filter (labeled LW)
which, along with the spectral response of the detector, acts as band pass for
longer wavelengths.

The narrow--band filter set includes Br$_\gamma$ and Fe II filters, plus the
associated K and H narrow-band filters to sample the  near--by continuum.
It is possible to insert in the beam several accurate grey filters to reduce
the flux from very bright sources.  Figure 2 shows the resolving power
(when associated with the $1\seco$ slit) and the efficiency of the available
grisms which are resin--replicated Milton-Roy gratings on IRGn6 prisms
(Vitali et al., 2000).

\begin{figure}
\centerline{\includegraphics[width=7.3cm]{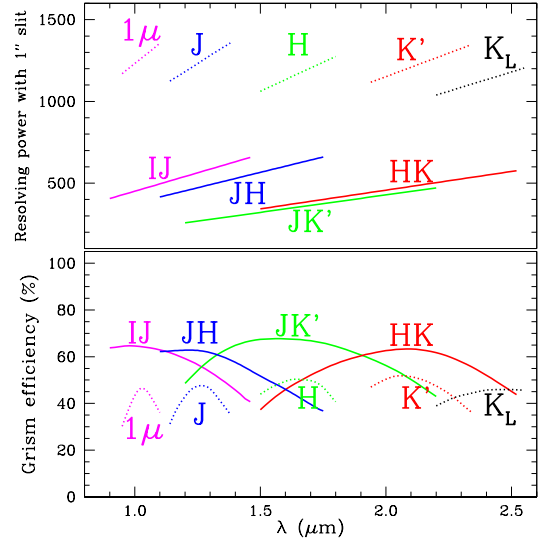}}
\caption{ 
Resolving power and efficiency of NICS grism.
}
\end{figure}

\section{General Description}
\label{opt}

The optical scheme comprises the single collimator and the two cameras;
all the optical components reside in a vacuum at a temperature of about
80 K, inside a suitable cryostat. The focal plane masks (field stops
and slits) are mounted on a wheel. Immediately after the focal plane,
a further removable field stop makes possible polarimetric measurements.

The collimator is an achromatic doublet lens, which images the entrance
pupil of the TNG and provides a parallel beam at the pupil plane
where Lyot stops can be placed.  Immediately after the pupil plane,
two adjacent wheels carry filters and grisms.  Two interchangeable optical
systems relay the image of the focal plane on the detector with
the desired magnification.

A third optical system images the entrance pupil on the detector, for
the purpose of an accurate alignment of the telescope with the instrument pupil,
and is not normally used in routine observations. A short discussion can be found in
Gennari et al. (1995).

The spectroscopic mode makes use of the wide--field camera by inserting one
of the grisms, located on the second filter wheel, into the parallel beam
after the pupil.

The rejection of stray and thermal light is left to the TNG baffles in
J, H, and K' bands.  At longer wavelengths, where the thermal radiation
could prevail, it is possible to insert a cold stop precisely positioned
at the pupil image.

The sensitive element of NICS is the Hawaii
 $1024 \times 1024$ pixels HgCdTe array detector 
(Rockwell Science Center).  It has a 18.5 $\mu$m/pixel
pitch and is sensitive to radiation at wavelengths between $\sim 0.90
\mu$m and $\sim 2.5 \mu$m.  Its performance in term of dark current,
efficiency, and read noise, is
comparable or better than the $256 \times 256$ NICMOS~3 (e.g. Lisi et al. 1996).

The electronic noise is dominated by the detector and by the first cold
amplifier and is $\sim 25~e^-$, if a suitable number of detector
resets (more than 32) is performed at each integration.
In the most common read strategy (double sampling),
during the integration at least two measurements are performed, one at
the beginning (to sample the reset bias level), the second at the end of
the integration ramp.

NICS is mounted at the same focal station as the optical CCD
camera; the two instruments share an adapter that carries
also the adaptive optics module. A plane mirror (M4), mounted on
a remotely--operated sliding bench, deflects the beam from the
telescope to the entrance of the IR camera, that has its optical axis
perpendicular respect to the telescope beam. NICS is mounted on the
adapter by means of a set of spherical joints that allows for a limited
adjustment of the optical axis alignment respect to the entrance beam. The
fine alignment is handled by the M4 mounting hardware, which is designed
to allow for small adjustments in matching the telescope optical beam to
the camera optical axis.

 \begin{figure}
 \vspace{-3.0cm}
 \centerline{\includegraphics[width=12.cm]{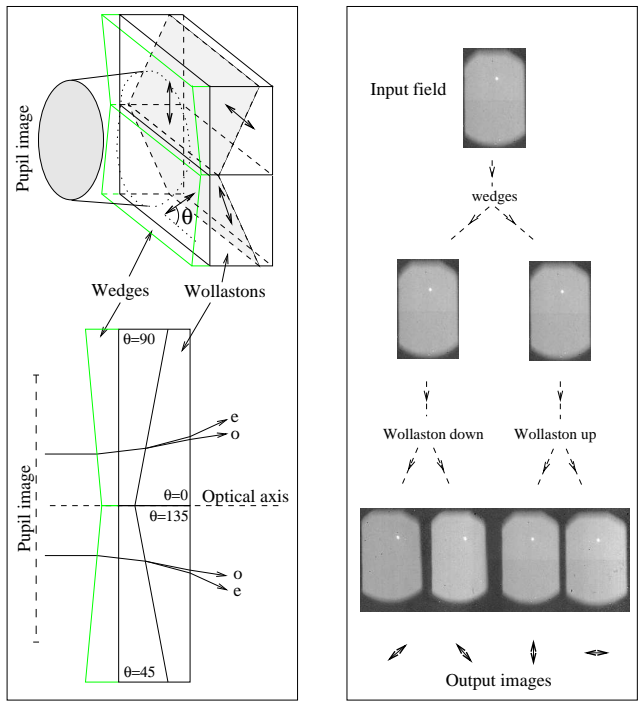}}
 \vspace{-3.0cm}
 \caption{ 
 Logical sketch of polarimetry operations. (Adapted from Oliva, 1997,
 but with real data.)
 }
 \end{figure}

\section{Optical Design}

The collimator used for all the scales and the observing modes is a
detached doublet (BaF$_2$--IRG2) (Oliva and Gennari 1998) with 
spherical-surface lenses that transform the f/11 TNG beam into a 22 mm parallel
beam (f/ larger than 2000) with a total level of aberrations below
0.4 mrad. The blur of the entrance pupil image has a maximum size of
0.21 mm at 90\% encircled energy, which guarantees an
accurate background rejection when the Lyot stop is inserted. The 
aberrations of the pupil image have been reduced by a factor 2 by
means of a mildly aspheric entrance window, which is designed to leave
the pupil plane position unchanged.

The two optical trains for the cameras and the pupil re--imaging system
are mounted on a wheel, accurately driven by an external motor. These
optics are based on detached doublets (BaF$_2$--IRG2) followed by one or
two lenses made of the same materials.

The wide-field camera (0.25"/pix) consists of four lenses. 
The spot blur
on the detector is good (90\% of the energy within a pixel inside a
$\approx 120\seco$ radius circle, 75\% in the corner). The exit pupil is
positioned 88 mm behind the focal plane of the camera,
which renders the system not completely
telecentric; this introduces some $cos^4$ losses (around $4.5\%$ at the
corners of the field).
In addition, the system suffers a moderate distortion of the
order of few percent at the field corners.

The small-field camera (0.13"/pix) consists of three spherical lenses.
Its overall image quality is much better than that of
the wide-field camera: the spot blur is smaller, distortion is basically
absent ($< 0.16\%$), and the exit pupil is at about -350 mm, which makes
the system almost telecentric.

The filters are mounted on two wheels located in the parallel beam just
after the pupil plane; the second wheel carries also the grisms, slightly tilted
with respect to the optical axis to achieve their best efficiency.  The
second filter wheel also hosts the wedged double Wollaston prism, made of
LiYF$_4$, that deviates the rays of the parallel beam into four different
beams corresponding to the polarization angles of 0, 45, 90 and 135
degrees; as a consequence, the input field of view is split into four
images on the detector, one for each of the four polarization angles (Oliva, 1997). This
way, one can perform simultaneous photometric measurements of the polarized
flux at different angles as required to derive the first
three elements of the Stokes vector (see figure 3).

For polarimetry, a suitable field stop right
after the focal plane limits the field of view to about 1/4 of the total
field, to avoid overlapping of the four polarized images on the detector
plane. The second wedged double Wollaston prism is made of LiNbO$_3$.
This prism is designed to perform
spectro-polarimetry when associated with one of the available grisms.
The polarizer works with the same principle as
the imaging mode, simultaneously delivering four polarized long-slit
spectra at angles rotated by 45 degrees. The $4\min$ slit normally used in
the spectroscopic mode will be masked to have an equivalent length of about
$50\seco$, to avoid any overlapping of the four polarized spectra.

In order to allow focusing capabilities, the detector is mounted
on a motorized base. For each optical configuration, the exact detector
position has been measured and the control software automatically
performs the internal focus setting.

\begin{figure}
\centerline{\includegraphics[width=20.0cm]{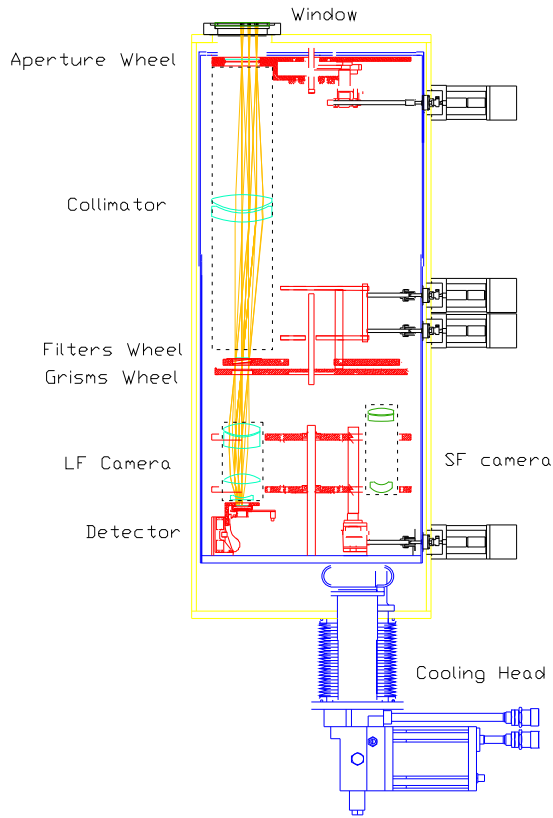}}
\caption{ 
Mechanical Layout of Nics cryostat.
}
\end{figure}

\section{Cryogenics and Mechanics}
\label{cry}

The cryo--mechanical design is based on a cylindrical vacuum shell, 
with a size dictated by the total length of the optical system and the
wheels diameter. This shape has the advantage of low weight and high
stiffness, associated with a reasonable cost.

The cylindrical case, with its symmetry axis orthogonal to the TNG exit
beam, is rigidly connected to the Nasmyth adapter flange 
and supports the internal cooled structure via thermally
isolating, rigid elements. The radiation shield is connected to the
cold structure; the volume between the pupil plane and the detector is
protected from stray light by means of a second radiation shield. The
internal structure supports all the optical functions, that is the four
wheels (mounted on ball bearings), the collimator, and the detector,
keeping them at the required relative positions.  Several external
micro--stepping motors take care of wheels and slides positioning,
as required by the selected observing mode.

Operations at the de-rotated Nasmyth focus necessitate the rotation of
the instrument itself, making impractical the use of liquid cryogenics
for cooling and maintaining the operative temperature; instead, we based
the cryogenic system on a closed cycle cooler.

Because of the relatively large mass of the instrument, cooling it down to
the operating temperature is best achieved by means of a continuous flow of
liquid nitrogen inside a pipe welded to the optical bench; during normal
operation the liquid nitrogen is not necessary.  The detector, in good thermal
contact with the cold plate, has an active temperature control.
An activated charcoal cryo--adsorption pump is installed on the
cold surface to guarantee a good vacuum in the presence of out-gassing
or small leaks for periods greater than 90 days. For maintenance purposes,
the molecular sieve is equipped with a suitable heating element and thermal
switch.

All measurements of temperature and pressure inside the cryostat are
transmitted to the control computer, where the software can look after
the regularity of operational parameters at scheduled intervals of time.

\section{Electronics}
\label{elect}

The data acquisition and control system is based on the controller
developed by the CCD Working Group of TNG, suitably modified to adapt it
to the architecture of infrared arrays (Comoretto et al. 1995, Comoretto
et al. 1999). The controller is based on a set of Transputer processors,
which are responsible for handling data and sending commands, and on a
DSP Motorola 56001, which generates the synchronized clock pattern needed
for accessing and reading the array multiplexer.

The analog signal read on each pixel of the four quadrants is buffered by
four FET amplifiers located on the same board that hosts the detector. 
After that stage, there is a 
set of four
16-bit A/D converters, which converts the pixel intensity of the four quadrants
in parallel.  
The parallel outputs from the converters are translated to the transputer
serial protocol using a dedicated programmable logic chip from Xilinx.

The transputer stage sends the digital data to a
Linux PC by means of a fiber link which exploit the fast serial
connection capability built into each Transputer. The controller takes
care of telemetry and stepper motors by means of dedicated RS--232
serial ports.

\section{Software}
\label{soft}

The low--level software involves relatively complex interactions between
the Motorola DSP and the cluster of Transputers modules, which operate
in a multitasking configuration, each performing an elementary task
in parallel. The transputer network is organized as a linear chain with the
possibility of single node branches (we call them {\sl left branches}), 
as described above.

Due to the intrinsic multitasking nature of transputers, we organized low
level software as a collection of modules, each performing an elementary task,
all acting in parallel. Complex tasks (as data acquisition, handling
and transfers) are realized by the cooperation of many modules, often
running on different CPUs.

One of the biggest issues we faced in developing transputer software was
the inter-processes communication. We developed a simple packet
switching solution, in a way that roughly resembles the IP protocol ({\sl Internet way}).
All communications are
performed by means of fixed--size packets. Each packet starts with a
header stating the node of origin, the destination node and 
destination process, and the command and sub-command(s). The packet has
also a large (1024 short integers) data area.

Each node has a process which examines the header of each packet and then
dispatches it to the three possible outward directions or to the
destination process for execution. The modified linear chain enables us
to make all routing by means of only {\sl local} fast comparisons, which makes the
system very efficient and suitable also for data dispatching.
For further details on the architecture of the interprocessor communication, see
Baffa et al. (1999), Baffa, (2000), and references therein.

The high--level control program, which comprise the telescope interaction, the
data handling and storing and the human interface of NICS (Xnics)
is based on a similar interface developed several years ago for the
Arnica IR camera.
Xnics provides the observer with the environment to define the parameters of
measurements and to start the desired sequence of integrations (such as single
frame, multiple frames, mosaics, scans, scripting language), along with
sky--source subtraction and preliminary reduction for quick--look purposes.
Several tools are available to control the overall quality of data during the
measurement, while the program is in charge of validating the
parameters which the observer has chosen before starting the observation. In
the background, a task is always active which monitors messages and error flags
coming from the low--level software.  

All the low level handling is performed by  a concurrent program, NICSgate. It
consists of several object-oriented processes, each one controlling a special
portion of the hardware functionality: telescope, motors, Transnix
initialization and programming, acquisition as a defined task and all types of
communication in real time. Each process maintains its inner state and can be
activated at any time, when an external event needs its special functions.

Due to the intrinsic {\sl network awareness} of both Xnics and NICSgate, a
distributed execution of the software is possible: NICSgate on the local
acquisition computer, Xnics on a remote one.

The software developed for this instrument is ``layer organized'', that is to
say organized as a stack of many layers of subroutines of similar levels of
complexity.  To accomplish its task, each routine needs relies only on the
immediately adjacent level and on global utility packages.  Such a structure
greatly simplifies the development and maintenance of the software.

Our efforts were aimed at several different requirements. Our first
priority was to have a flexible laboratory and telescope engine, capable of
acquiring easily the large quantity of data a panoramic IR array can produce.
Another main goal was to produce an easy-to-use software with the
smallest ``learning curve''. Our idea was that data acquisition must {\sl
disappear} from observer attention, giving him/her the possibility to
concentrate on the details of the observations; in this way, observing
efficiency is much higher. The human interface is realized through a fast 
menu interface. The operator is
presented only with the options which are currently selectable, and the menu is
rearranged on the basis of user choices or operations.  We have also
implemented automatic procedures such as multi--position (``mosaic''),
multi--exposure (stack of many frames) and a {\sl scripting language} capable
of performing a fairly complex set of measurements with only a ``quality
control'' from the observer.

\section{Characterization and performance}
\label{detec}

NICS was tested and characterized both in the laboratory and, at the
moment of writing, during four commissioning runs at the TNG.

\begin{table}
\begin{center}
\begin{tabular}{lccc}
Filter        & J$_S$ & H & K' \\
\hline
Zero point    & 22.1  & 22.3  & 21.8  \\
Efficiency    & 0.21  & 0.28  & 0.32  \\
Lim. mag.     & 22.5  & 21.4  & 21.2  \\
Average sky   & 15.5  & 13.2  & 13.2  \\ 
mag/arcsec$^2$&       &       &       \\
\hline
\end{tabular}
\end{center}
\caption{Zero points, efficiencies and limiting magnitudes of NICS at the
TNG measured in October-December 2000. 
Lim. Mag. is the point source limiting magnitude
for a 3$\sigma$ detection in a hour of on-source integration
with a seeing of 1$''$ (integration aperture of 2$''$) in LF mode. 
Average sky is the approximate sky brightness, in magnitudes/arcsec$^2$ 
of the various measurements.
}
\end{table}

For what concerns the detector, the current read-out noise (double sampling)
is about 25 e$^-$, but is expected to improve once the forthcoming
multi-sampling mode will be available. The dark current is about 1 e$^-$/sec
and the well capacity is about 10$^5$e$^-$. With the current electronics
setup the conversion factor is about 8 e$^-$/ADU.

The efficiency of the instrument was measured in the laboratory and at the telescope.
In the laboratory, 2$\mu$m efficiency was measured by means
of a blackbody located in front of the window. The efficiency in imaging
mode is about 60\%, in agreement with what is expected from the combination
of the efficiencies of the detector and of the filters.
The overall efficiency of the instrument and telescope (and atmosphere)
was measured during the commissioning runs. The total efficiency critically
depends on the cleanness of all optical surfaces, which, for various reasons 
could be less than perfect.
The efficiencies and sensitivities given
in the following, refer mostly to the last two commissioning runs, after the
cleaning of M3 (the relay mirror to the Nasmyth focus), but without cleaning
of the entrance window. The {\sl final} assessment of NICS' performance
at the TNG must await the cleaning of all optical surfaces.
In Table 1 we give the zero points in the three main broad bands
and the corresponding (total) efficiencies. We report also the limiting
magnitudes for a detection at 3$\sigma$ in a hour of on-source integration
with a seeing of 1$''$ (integration aperture of 2$''$). 

\begin{figure}
\centering
\includegraphics[width=0.8\linewidth]{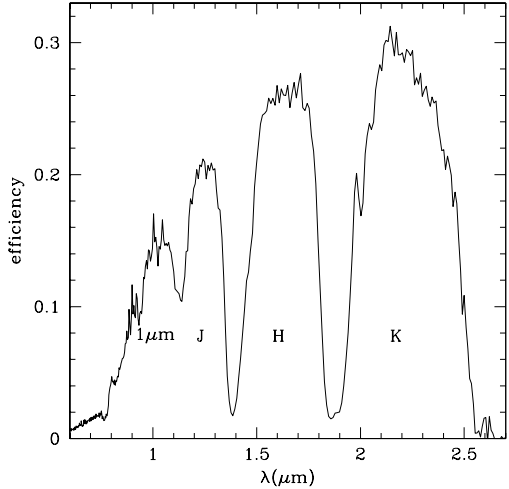}
\caption{Global efficiency of the system (instrument + telescope +
atmosphere) measured through the Amici prism.}
\end{figure}

The efficiency in the spectroscopic modes agrees with the efficiencies
given in Table~1 convolved with the response curves given in Fig.~2. 
Particularly interesting is the efficiency measured through the Amici prism,
since the latter disperser has a nearly flat efficiency over the whole
near-IR range and, therefore, gives a global view of the efficiency of the
system. Fig.~5 gives the ({\it absolute})
efficiency obtained through the Amici and,
more specifically, by dividing the Amici spectrum of a standard star by
the intrinsic spectrum of the star (therefore, this is global efficiency
of the Amici prism + instrument + telescope + atmosphere). Note the main
atmospheric windows which are marked in the figure. Also worth noting
is the efficiency drop in the J and 1$\mu$m bands 
which can be ascribed to a drop in efficiency of the detector at short wavelengths.

\begin{table}
\begin{center}
\begin{tabular}{ccc}
Disperser     & Central Resolution & Limiting Magnitude \\
\hline
Amici         & 50             & 19.8           \\
\hline
JK'           & 350            & 18.4           \\
\hline
IJ            & 500            & 18.6           \\
JH            & 500            & 18.3           \\
HK            & 500            & 18.1           \\
\hline
I             & 1250           & 17.8           \\
J             & 1200           & 17.6           \\
H             & 1150           & 17.6           \\
K             & 1250           & 17.2           \\
\hline
\end{tabular}
\end{center}
\caption{Limiting magnitudes of NICS at the
TNG in different spectroscopic mode.
Limiting Magnitude is the point source limiting magnitude
for a 3$\sigma$ detection in a hour of on-source integration
with a seeing of 1$''$ and with a 1$''$ slit in LF mode,
averaged on the wavelength range. 
The sky luminosity is assumed to be the same as in Table~1.
}
\end{table}

A broad estimate of the limiting magnitude in polarimetry can be derived from 
broad--band values by subtracting 1.5 magnitudes (pupil light is divided in four parts).
In Table~2 we give preliminary values for spectroscopic 
limiting magnitudes using large fields, one hour of exposure, and with a 1$''$ slit.

As mentioned in Section 4, the wide-field camera optics suffer from
distortion at the level of a few percent at the array vertices.  During
commissioning, this distortion was characterized at the telescope by
measuring crowded stellar fields with known astrometry.  It turns out
that the distortion can be well approximated by a symmetric radial
sixth-order polynomial, and the coefficients for the forward and
inverse transformations were derived from astrometric
measurements.  Results show that, during commissioning, the optical
center of the array is within one pixel of the center of symmetry of
the distortion, and the amplitude of the measured distortion is
consistent with, perhaps slightly smaller than, the design
specifications.

\section{Conclusions }
\label{conclusions}

After several months of testing at TNG, NICS proved able to provide
the entire set of observing modes included in the design with the desired
performance. At present, the TNG is equipped with a near-IR facility
well suited to a 3.5m-class telescope, ready to serve the astronomical
community. 

Two of the available observing modes, polarimetry and low resolution
simultaneous 0.9--2.5 $\mu$m spectroscopy are unique to NICS and make
the TNG + NICS system the only facility available for this kind of
observations.

\bigskip\bigskip

\begin{acknowledgements}
Most of the instrumentation projects for astronomy can be successfully
achieved only with the help of a team of skilled engineers and
astronomers. The authors would like to emphasize that the NICS project received
(and is still receiving) support, help and technical input from several people;
they thank in particular 
A. Marconi, L. Miglietta, G., Tofani, F. Fusi Pecci,
L. Corcione, G. Nicolini, J. Licandro, 
the TNG development team, the CCD Group, and many people from Arcetri 
and from the astronomical community for useful discussions.
The authors would like to particularly acknowledge the support coming 
from the TNG
staff, always ready to help during the difficult and busy time of the
commissioning at the telescope.
\end{acknowledgements}

\section*{References}

\myitem
Barbieri C., TNG Technical Report n.~53, November 1995

\myitem
Baffa C., and Comoretto G., 1996, Arcetri Technical Report N.3/96

\myitem
Baffa C., Comoretto G., Gavrioussev, V., Giani, E., Lisi F., 1999, 
{\em Mem.\ Soc.\ Astron.\ It.}  in~press  

\myitem
Baffa C., 2000, Arcetri Technical Report N.2/2000

\myitem
Comoretto G., Baffa, C. Gavrioussev, V., Giani, E., Lisi F., 1999, 
{\em Mem.\ Soc.\ Astron.\ It.}  in~press  

\myitem
Comoretto G., Baffa C., and Lisi F., 1995, Arcetri Technical
Report N.4/95

\myitem
Fusi Pecci F., Stirpe G.M., TNG Instrument Plan, 1992, Osservatorio
Astronomico di Bologna.

\myitem
Fusi Pecci F., Stirpe G.M., Zitelli V., TNG Instrument Plan:II.
A Progress Report,  1994, Osservatorio Astronomico di Bologna.

\myitem
Lisi F., Baffa C., Biliotti V., Bonaccini D., Del Vecchio C.,
Gennari S., Hunt L.K., Marcucci G., Stanga S., 1996, \PASP{108}{p364}.

\myitem
Gennari S., Vanzi, L., Lisi F., 1995, \SPIE{2475}{221}.

\myitem
Maiolino R., Rieke G.H., Rieke M.J., 1996, \AJ{111}{537}

\myitem
Oliva E., 1997, \AAS{123}{589}

\myitem
Oliva E., Gennari S., 1998, \AAS{128}{5890}

\myitem
Oliva E., Origlia L. 1992, \AaA{254}{466}

\myitem
Ragazzoni R.  (ed.), AdOpt Yearly Status Report, 1996, Osservatorio Astronomico di Padova

\myitem
Vitali, F., Cianci, E., Lorenzetti, D., Foglietti, V., Notargiacomo, A., Giovine, E., Oliva, E., 
2000, \SPIE{4008}{1383}.

\end{document}